\documentclass{PoS}
\newcommand{\be}{\begin{eqnarray}}
\newcommand{\ee}{\end{eqnarray}}
\newcommand{\ta}{\tau_a}

\title{Toward precision jet event shape for future Electron-Ion Collider}

\ShortTitle{Toward precision jet event shape for future Electron-Ion Collider}

\author{Daekyoung Kang \thanks{This work  is supported by the National Natural Science Foundation of China (NSFC) through Grant No. 11875112 and by the China Postdoctoral Science Foundation through Grant No.KLH1512104}\\
        Key Laboratory of Nuclear Physics and Ion-beam Application (MOE) and Institute of Modern Physics, Fudan University, Shanghai 200433, China\\
        E-mail: \email{dkang@fudan.edu.cn}}

\author{\speaker{Tanmay Maji}\\
Key Laboratory of Nuclear Physics and Ion-beam Application (MOE) and Institute of Modern Physics, Fudan University, Shanghai 200433, China\\
        E-mail: \email{tanmay@fudan.edu.cn}}

\abstract{ We present angularity differential cross-section for the deep inelastic scattering process (DIS) in the framework of soft-collinear effective theory (SCET). Using SCET, the cross-section is factorized in terms of hard, jet, beam and soft functions. Our result includes resummation of the large logarithms up to next-to-next leading logarithmic (NNLL) accuracy. The numerical results presented here for DIS angularity cross-section can be explored by the future EIC.
}

\FullConference{LC2019-QCD on the Light-Cone from hadron to heavy Ions\\
		16-20 September 2019\\
		Ecole Polytechnique, France}

\begin{document}
\section{Introduction\label{intro}}

In high energy scattering, the most common final states are collimated branches of strongly interacting particles, called jet, and the jets are characterized by classic QCD variables called event shapes. Event shapes characterize the geometry of final states in a jet like structure and plays a significant role to improve our understanding on the basic aspects of QCD in high-energy scattering, such as short distance phenomenon in the underline events, initial and final state radiation as well as non perturbative effects. 
This includes the precise  determination of strong coupling $\alpha_s$.  
There are several event shapes which carries the substructures of jets e.g., two-jet event shapes are thrust, jet masses, jet broadening and later on event shape angularity is introduced. Angularity event shape $(\ta)$ is introduced in \cite{Berger:2003iw} and has drowned a special attention as it depends on a continuous parameter $a$ unlike other event shapes. 
The limiting case of angularity event shape relates with the other event shapes measured experimentally e.g., at $a = 0$, angularity reduces to $thrust$  and at $a = 1$, it reduces to $jet~ broadening$. 

The event shapes have been measured to high accuracy in several experiments e.g., $ e^+e^- $ event shape with center-of-mass energies $Q=91.2$ GeV and $Q=197$ GeV are measured at LEP \cite{Achard:2011zz}, DIS event shapes are measured in HERA by the ZEUS and H1 collaborations \cite{Adloff:1997gq,Adloff:1999gn,Aktas:2005tz,Breitweg:1997ug,Chekanov:2002xk,Chekanov:2006hv}. In the recent future Electron-Ion collider (EIC) experiment \cite{Accardi:2012qut} , proton will be probed  more extensively by high energy lepton and could be considered measuring more general event shapes angularity. 

Theoretically, Soft-collinear effective theory (SCET) \cite{Bauer:2000ew,Bauer:2000yr,Bauer:2001yt,Beneke:2002ph} provides a systematic way to achieve high precision in the high energy scattering and has become a powerful tool to study the event shapes.  The SCET has been used to provide precision prediction to  $e^+e^-$ event shapes extensively e.g., SCET has been used to achieve $N^3LL$ resummation of thrust \cite{Becher:2008cf} and heavy jet mass \cite{Chien:2010kc} in $e^+e^-$ collision. Angularity in $e^+e^-$ is presented at NNLL accuracy in \cite{Bell:2018gce}. SCET has also been used to predict the event shapes in $pp$ collisions \cite{Stewart:2010pd,Stewart:2010tn,Berger:2010xi,Jouttenus:2013hs}. Whereas, very few attempts to achieve high precision in deep inelastic scattering (DIS) event shape predictions using SCET e.g., DIS thrust is studied in \cite{Kang:2013nha,Kang:2014qba}, where the thrust is computed for a particular axis choice along the jet axis and considered contribution from both hemispheres. 

 Here in this paper, we present the angularity event shapes in deep inelastic scattering (DIS) process in the framework of SCET and give NNLL prediction to the future EIC measurement.

\section{Angularity in DIS}\label{ang}
In DIS, an electron with momentum $k$ scatters off a proton of momentum $P$ by exchanging a virtual photon with a large momentum transfer $ q $. The space-like photon momentum can be written by a positive definite quantity $Q^2=-q^2$, where $ Q$ sets the momentum scale of the scattering $Q \gg \Lambda_{QCD}$.   Bjorken scaling variable $x=\frac{Q^2}{2 P.q}$ ranges between $0 \leq x \leq 1$ and inelasticity  $y=\frac{Q^2}{x s}$ ranges between $0 \leq y \leq 1$ .  

Angularity for DIS is defined by 
\be
\ta = \frac{1}{Q} \sum_i |p^i_\perp| e^{-|\eta_i|(1-a)},
\ee
where the sum over all final state particles $i$ with rapidity $\eta_i$ transverse momentum  $p^i_\perp$. The DIS angularity can be expressed in terms of the four-vectors $q_B$ 
and $q_J$  
as given by
\be
\ta = \frac{2}{Q^2} \sum_{i \epsilon \mathcal{X}} min \bigg{\{}(q_B.p_i) \bigg( \frac{q_B.p_i}{q_J.p_i} \bigg)^{-a/2} ,~ (q_J.p_i)\bigg(\frac{q_J.p_i}{q_B.p_i} \bigg)^{-a/2}\bigg{\}}, \label{tau2}
\ee 
where $\mathcal{X}$ stands for $\mathcal{H}_B, \mathcal{H}_J$ for beam hemisphere and jet hemisphere respectively. Transverse momentum $p^i_\perp$  is defined with respect to corresponding $q_J$ and $q_B$.  We can have different choices for the reference vectors $q_{J}$ and $q_B$. For example-- 
choice-I, $\ta^{(1)}$:  $q_B=xP$ (along the beam axis) and $ q_J$ is along the physical jet axis. Choice-II, $\ta^{(2)}$:  $q_B=xP$  and $ q_J=q+xP$. Measurement of $\ta^{(2)} $ groups final state particles into back-to-back hemisphere in the Breit frame. One needs to prove the SCET factorization theorem for different axis choice separately as the final state radiations in DIS are probed differently for these variables $ \ta^{(1)}$ and  $ \ta^{(2)}$. Also, each choice has different sensitivity to the transverse momentum. Here we consider the choice-I, where $q_J$ align to final jet axis. 


\section{Factorization}\label{fact}
In the theoretical description of event shapes distributions, according to the factorization theorem, the cross section can be computed through a product of probability distribution functions, namely, the jet function $(J_q)$ that describes collinear emissions of the final state radiation into the jet direction, the beam function $(B_q)$ that created from initial state radiations  encodes the information about partonic structure of the incoming hadron, the soft function $(S)$  that created from the soft radiation and the hard functions $(H)$. For generic values of the event shape, the distribution can be described in fixed-order perturbation theory.  The perturbative expansion  develops large logarithmic corrections that need to  be resummed to all orders. Resummation of large logs for a event shape distributions have been pushed to new level of accuracy with methods from SCET.  Under the soft-collinear factorization the DIS angularity cross section can be expressed as
\be \label{eq:fact}
\frac{d\sigma}{dx\, dQ^2d\, \ta} =& \frac{ d\sigma_0}{dx dQ^2}  \sum_\nu H_\nu (Q^2,\mu)  & \int d\ta^J \, d\ta^B\, dk_S \ J_q(\ta^J,\mu)  B_{
\nu/q}(\ta^B,x, \mu) \nonumber\\
&&\times   S(k_S,\mu)  \delta\Bigl(\ta-\ta^J-\ta^B-\frac{k_S}{Q_R} \Bigr),
\ee
where, $\nu$ runs over quarks and antiquark flavors.  $H_\nu$ is the hard function arises integrating out the hard degree of freedom from QCD in matching on to SCET and $J_q,B_q$ and $s$ are the jet, beam and soft functions for angularity observable. The hard, soft, jet and beam functions are defined at their canonical scales at which the logarithms are minimized $\mu_H = Q,  \mu_S = Q \ta, \mu_{J,B} = Q \ta^{1/(2-a)}$ and evolved to a desire scale $\mu$ governed by the solution of the renormalization group equation (RGE). 

The whole cross-section can be characterized in three physical regions: the peak region ($\tau_a \sim 2 \Lambda_{QCD}/Q \ll 1$), the tail region ($  2 \Lambda_{QCD}/Q \ll \tau_a  \ll 1$) and the far-tail region  ($ \tau_a \sim 1$).  For the peak region $(\ta \ll 1)$, the non-perturbative effect dominates. In the tail region the distribution functions $(J_q,B_q,S)$ are at the canonical scales for which the logs in the fixed-order functions are minimized. The evolution from canonical scale to another scale $\mu$ resums the logs of the ratio $\mu/\mu_{J,B,S}$ to all order in $\alpha_s$. In the SCET approach we stop the renormalization group evolution well above $\Lambda_{QCD} $. In the far-tail region, at the large values of $\ta$, the logs are no longer large and the non-singular term becomes equally important and one need to tern off the resummation. This property for different region can be achieved by using $profile ~ functions$ \cite{Bell:2018gce}. 
 One-loop Jet and soft functions are calculated analytically in \cite{Hornig:2009vb}. Two-loop numerical results are presented in \cite{Bell:2018vaa} for soft function and in \cite{Bell:2018gce} for jet function. The angularity beam function for DIS at the $\mathcal{O}(\alpha_s)$ is first computed in this work.

\section{Result and discussions}\label{rlt}
\begin{figure}[h]
\begin{center}
\includegraphics[scale=0.25]{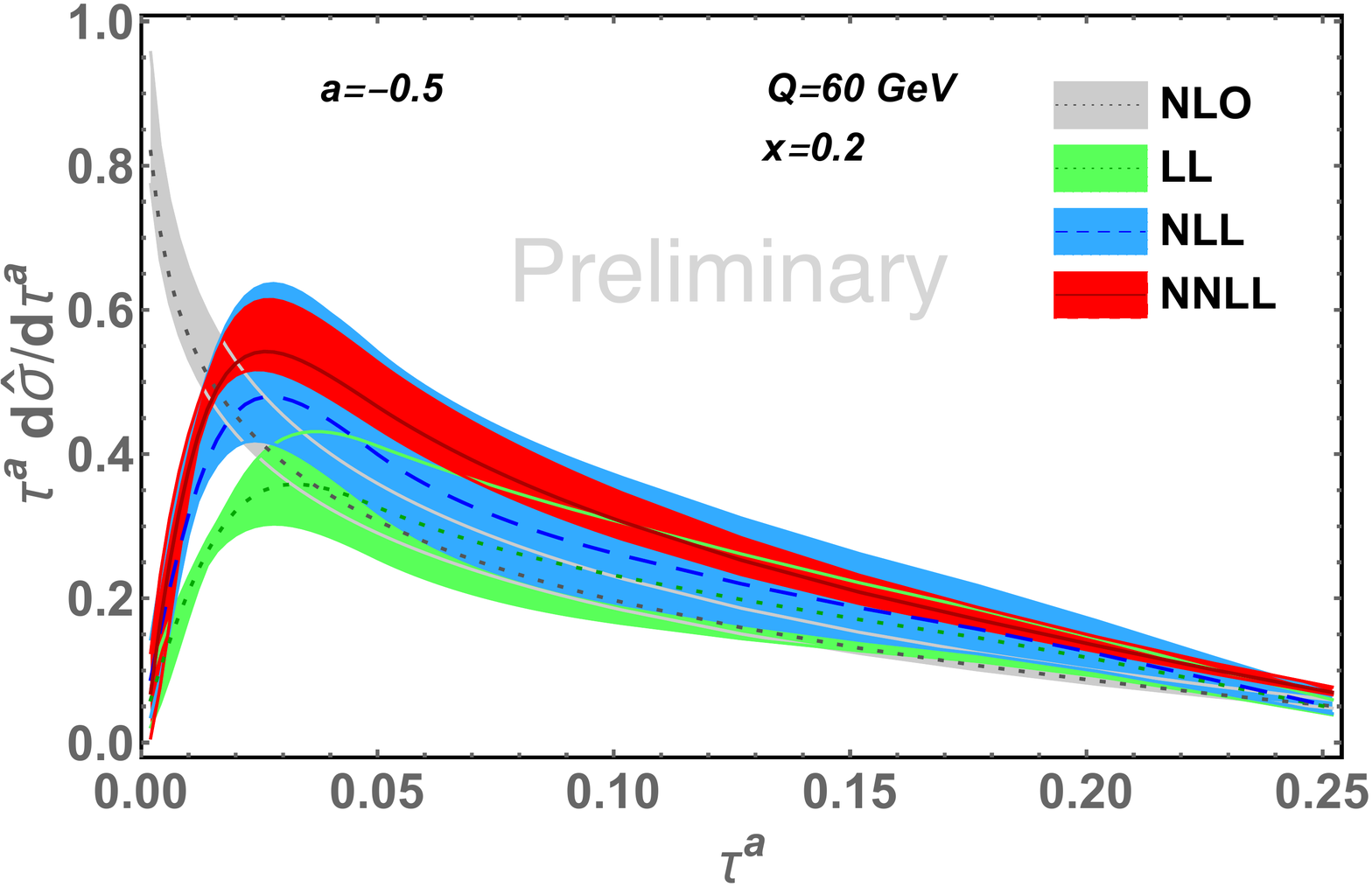} 
\includegraphics[scale=0.25]{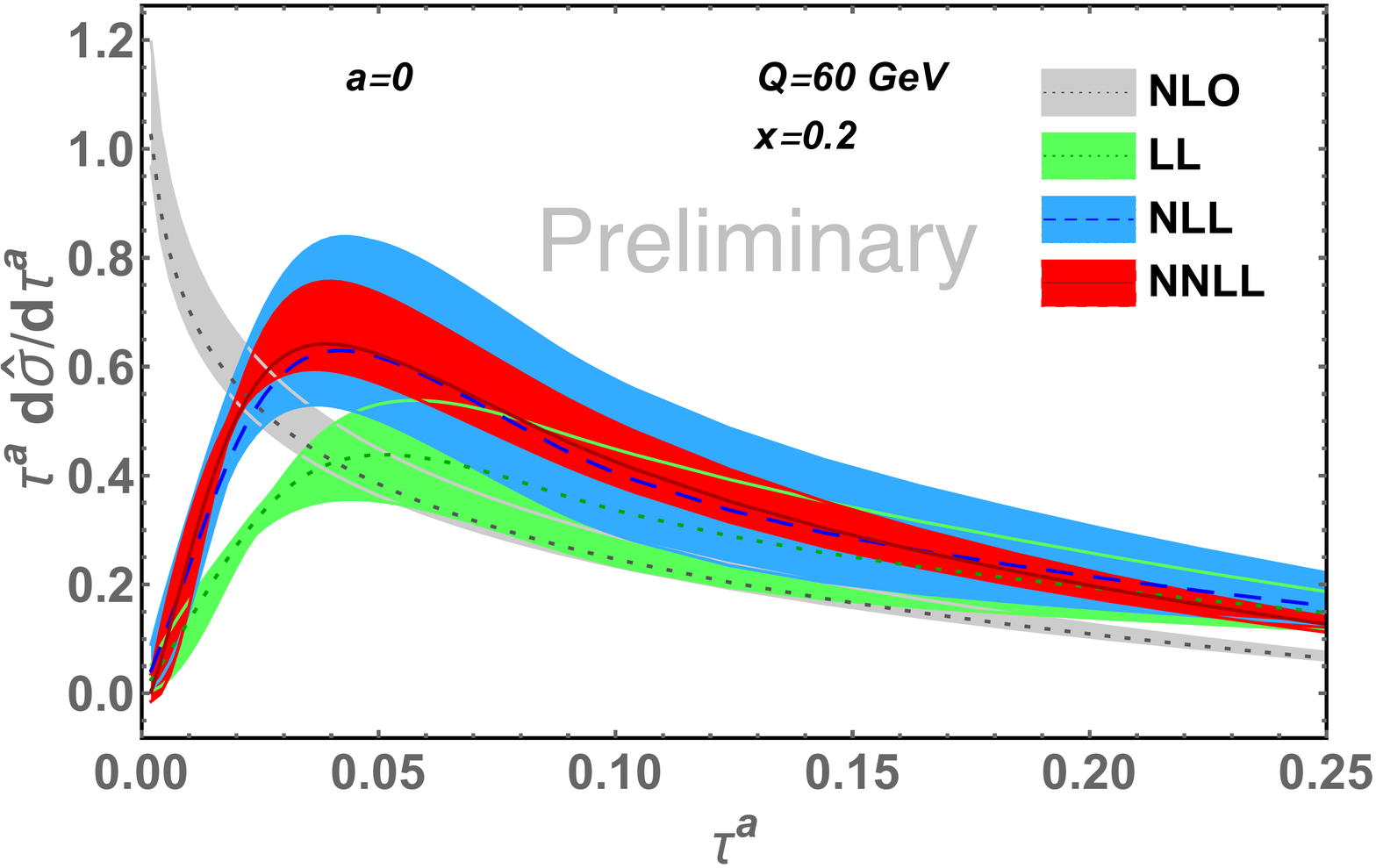} 
\caption{\label{dsig} Preliminary results of weighted differential cross-section for DIS angularity  $
\tau_a \frac{d\hat{\sigma}}{d \tau_a}$ is shown at different $a=-0.5~(left)$, $0 ~(right)$ and $\sqrt{s}=140$ GeV. The band indicates perturbative uncertainty.} 
\end{center} 
\end{figure}

We compute the one-loop beam function generated by the one gluon emission in the initial state radiation. The beam function can be expressed as a convolution of parton distribution function (PDF) and a coefficient at  $\mathcal{O}(\alpha_s)$. Using this one-loop beam function and the other functions from \cite{Hornig:2009vb,Bell:2018gce} in Eq.(\ref{eq:fact}), we calculate the differential cross section at $\mathcal{O}(\alpha_s)$. The precision in the cross section is governed by the resummation of the large logs in the individual distribution functions. To demonstrate the improvement in precision prediction in the differential cross-section, we present the LL, NLL and NNLL results along with the NLO results for the EIC kinematics in fig.\ref{dsig}. The EIC will cover a wide longitudinal momentum fraction range and rapidity range  $0.01\leq y \leq 0.95$ for the center of mass energy $\sqrt{s}=45$ and $140~ GeV$ \cite{Accardi:2012qut}.

Fig.\ref{dsig} shows weighted differential cross-sections $\tau_a \frac{d\hat{\sigma}}{d \tau_a} =\tau_a \frac{1}{\sigma_0}\frac{d\sigma}{d \tau_a} $  for different values of angularity parameter $a=-0.5$ and $0$ at a fixed $x=0.2$ and $Q=60~ GeV $. As the differential cross-section falls vary rapidly, we multiply by a wait factor $\tau_a$ for better visibility. We choose $\sqrt{s} = 140~GeV$ as EIC plans to achieve $\sqrt{s}=45$ and $140~ GeV$.  For the parton distribution function in the beam function, we use the MSTW2008 set and consider five quarks and antiquarks excluding top quark. 
Each sub-figure contains four plots: the NLO (singular) result with perturbative uncertainty is illustrated in gray and the resummed results at the LL, NLL and NNLL accuracy are shown in color band green, blue and red respectively. The difference between the NLO and NNLL results shows the effect of the resummation. In the peak region, NLO result blows up as $\ln(\tau_a)/\tau_a$, while the NLL, NNLL results converge due to the resummation of the large logs. Whereas, in the tail region, the resummation effect is small and a less significant deviation leads to the higher precision. In this tail region the non-singular terms become significant. The bands in respective colors represent the perturbative uncertainty generated by varying the scales $\mu_{H,S,J,B}$ given by the "profile functions" presented in \cite{Bell:2018gce}. The uncertainty band in the peak region is dominated by the soft scale variation.
The sub-figures in fig.\ref{dsig} indicate that the peak moves to right with the increasing value of $a$ and also the peak value increases. The results for $a=0$ represents the differential cross-section for DIS thrust corresponding to this kinematics.

\section{Conclusions}\label{con} 
The future EIC will cover the kinematics  $0.01\leq y \leq 0.95$ for $\sqrt{s}=45, ~140~ GeV$ and our choice of the $x-value$ and $Q-value$ could be one of the choices that EIC may measure at. We show the numerical results for DIS angularity cross-section at the different values of angularity parameter $a=-0.5$ and $0$.  Result includes the precision upto NNLL accuracy and give prediction to the future EIC. 



\end{document}